\documentclass[conference]{IEEEtran}
\IEEEoverridecommandlockouts
\usepackage{cite}
\usepackage{amsmath,amssymb,amsfonts}
\usepackage{graphicx}
\usepackage{textcomp}
\usepackage{xcolor}
\usepackage{afterpage}

\usepackage{amsmath}
\usepackage{tikz}

\usepackage{graphicx} 
\usepackage{cellspace} 
\usepackage{array} 
\usepackage{xcolor} 
\usepackage{graphicx} 

\usepackage{algorithm}
\usepackage{algpseudocode}
\usepackage{algorithm}
\usepackage{float} 
\usepackage{algpseudocode}
\usepackage{subcaption}
\usepackage{graphicx}
\usepackage{array}
\usepackage{balance}

\usepackage{amsmath}
\usepackage{todonotes}
\usepackage[hidelinks]{hyperref} 

\usepackage{booktabs}
\usepackage{caption}

\usepackage{placeins}
\usepackage{hyperref}
\usepackage{xcolor} 

\usepackage{multirow} 
\usepackage{graphicx} 
\usepackage{tabularx}

\hypersetup{
    colorlinks=true, 
    linkcolor=blue,  
    citecolor=blue,  
    urlcolor=blue    
}

\usepackage{fancyhdr}

\fancypagestyle{IEEEtitlepage}{%
    \fancyhf{} 
    \fancyhead[C]{\footnotesize THIS PAPER HAS BEEN ACCEPTED AT THE 12th Annual Conference on Computational Science \& Computational Intelligence (CSCI’25) AND FINAL VERSION IS SUBMITTED}
    \fancyfoot[C]{\footnotesize
    The final published paper should be cited as: Amr Akmal Abouelmagd, Amr E Hilal, "Leveraging the Power of AI and Social Interactions to Restore Trust in Public Polls," The 12th International Conference on Computational Science \& Computational Intelligence (CSCI), Las Vegas, Dec 3, 2025}

}

\fancypagestyle{IEEEotherpage}{%
    \fancyhf{}
    \fancyhead[C]{\footnotesize THIS PAPER HAS BEEN ACCEPTED AT THE 12\textsuperscript{th} Annual Conference on Computational Science \& Computational Intelligence (CSCI’25) CONFERENCE AND FINAL VERSION IS NOT PUBLISHED YET.}
    
}

\makeatletter
\let\ps@IEEEtitlepagestyle\ps@IEEEtitlepage
\makeatother

\def\BibTeX{{\rm B\kern-.05em{\sc i\kern-.025em b}\kern-.08em
    T\kern-.1667em\lower.7ex\hbox{E}\kern-.125emX}}
\begin{document}

\title{Leveraging the Power of AI and Social Interactions to Restore Trust in Public Polls\\

}

\author{\IEEEauthorblockN{Amr Akmal Abouelmagd}
\IEEEauthorblockA{\textit{Department of Computer Science} \\
\textit{Tennessee Technological University}\\
Cookeville, TN, USA \\
aabouelma42@tntech.edu}
\and
\IEEEauthorblockN{Amr Hilal}
\IEEEauthorblockA{\textit{Department of Computer Science} \\
\textit{Tennessee Technological University}\\
Cookeville, TN, USA \\
ahilal@tntech.edu}
}

\maketitle

\thispagestyle{IEEEtitlepage}
\pagestyle{IEEEotherpage}

\begin{abstract}

The emergence of crowdsourced data has significantly reshaped social science, enabling extensive exploration of collective human actions, viewpoints, and societal dynamics. However, ensuring safe, fair, and reliable participation remains a persistent challenge. Traditional polling methods have seen a notable decline in engagement over recent decades, raising concerns about the credibility of collected data. Meanwhile, social and peer-to-peer networks have become increasingly widespread, but data from these platforms can suffer from credibility issues due to fraudulent or ineligible participation. In this paper, we explore how social interactions can help restore credibility in crowdsourced data collected over social networks. We present an empirical study to detect ineligible participation in a polling task through AI-based graph analysis of social interactions among imperfect participants composed of honest and dishonest actors. Our approach focuses solely on the structure of social interaction graphs, without relying on the content being shared. We simulate different levels and types of dishonest behavior among participants who attempt to propagate the task within their social networks. We conduct experiments on real-world social network datasets, using different eligibility criteria and modeling diverse participation patterns. Although structural differences in social interaction graphs introduce some performance variability, our study achieves promising results in detecting ineligibility across diverse social and behavioral profiles, with accuracy exceeding 90\% in some configurations.


\end{abstract}

\begin{IEEEkeywords}
Social Computing, Centrality Measures, Social Network Analysis, Social Interaction, Public Polls
\end{IEEEkeywords}

\section{Introduction}
Crowdsourced data has transformed social science research by providing large-scale, real-time insights into human behavior, enabling researchers to analyze public opinion, social movements, and cultural trends more efficiently \cite{salganik2019bit}. By leveraging the wisdom of the crowd, social scientists can enhance the diversity and representativeness of data, offering new opportunities for studying complex social phenomena such as collective decision-making and public policy \cite{howe2009crowdsourcing}. Crowdsourced data, like public opinion surveys or public polls, often suffers from nonresponse bias, where certain demographic groups are underrepresented, potentially skewing results and limiting the generalizability of findings \cite{kohut2012assessing, keeter2023publicopinion}. Moreover, there is a growing difficulty reaching and persuading potential respondents \cite{kohut2012assessing}. For example, the telephone survey response rates declined from 36\% in 1997 to 3\% in 2023 \cite{keeter2023publicopinion}. On the other hand, there is an observed growth in online opt-in polling compared to conventional methods like phone surveys. While promising, the participation rate in the online polls, and consequently the error rate, are still concerns as it might mislead the public as was evidenced by the significant errors in the 2016 and 2020 US presidential election polls \cite{kennedy2018evaluation}, \cite{uselection2024}. A post-election review of polling by the American Association for Public Opinion Research (AAPOR) cited ``polling error of an unusual magnitude'' in the 2020 polls \cite{kennedy2023public}. Recurring errors in influential public polls highlight the need for new polling paradigms.

Decentralized peer-to-peer polling can expand participation in surveys and decision-making, but credible outcomes require strict eligibility control. In systems like e-voting \cite{gibson2016review, Ohize2024}, eligibility is usually enforced by a central authority, which weakens decentralization and adds friction, especially for polling. Removing this dependency simplifies participation and encourages adoption.

To address this problem, we explore how trust in polling data can be restored by leveraging social interactions among participants via social networks. Today, social networks serve as conduits for information, often accompanied by virtual interactions like comments, shares, and likes. We hypothesize that observing social interactions with sufficient dissemination flows can reveal behavioral patterns to infer participants' eligibility without a central authority or relying on people's honesty. We assume honesty as a common operational foundation in our study and investigate the effect of changing its ratio among the participants. This is similar to the assumption that underlies cryptocurrency systems \cite{nakamoto2008peer, 9299581}, which require at least two-thirds of participants to be honest for proper functioning \cite{lamport2019byzantine}. By combining social behavior with decentralized data-sharing, we aim to develop an AI-based polling mechanism that ensures broad participation and reliability.


In this paper, we explore this concept by addressing the following research questions: (1) How can polling be effectively structured within a practical model based on social dissemination? (2) To what extent can social interactions among participants be observed and analyzed to establish trust in the collected data? (3) What social and technical factors influence the success of this process?

\textbf{Contributions.} To answer these research questions, we present an empirical study aimed at detecting ineligible participation in a polling environment that rely on peer-to-peer communication. We propose a mechanism in which a dissemination graph is constructed using the flow of sharing a polling requests among peers. We leverage centrality measures and machine learning applied to the dissemination graph data to identify ineligible participation. Our contributions can be summarized as follow:

\begin{itemize}

    \item We propose a data dissemination mechanism for a peer-to-peer polling system that leverages socially induced interactions of poll sharing to learn participants' eligibility. The mechanism encourages both active participation and continued propagation of information.
    

    \item We simulate the proposed mechanism using real-world datasets, incorporating different social interaction patterns and eligibility criteria that influence information dissemination in peer-to-peer settings. A dissemination graph is constructed by observing the social dynamics underlying data-sharing activities.

    \item We employ Graph Neural Networks (GNNs) to analyze the dissemination graph and assess participant eligibility, demonstrating prediction accuracy across various factors that influence the effectiveness of the proposed mechanism.
 


\end{itemize}

The remainder of this paper is organized as follows. Section~\ref{sec:2} reviews related work. Section~\ref{sec:3} introduces the proposed system and key definitions underlying our approach. Section~\ref{sec:4} details the system architecture and dynamics, including node types, participation mechanisms, and data forwarding. Section~\ref{sec:5} presents the empirical evaluation with the experimental setup and performance analysis. Section~\ref{sec:6} reports the main results and observations. Section~\ref{sec:7} discusses the study’s contributions and limitations. Finally, Section~\ref{sec:8} concludes and outlines directions for future work.


\section{Related Work} \label{sec:2}
Our proposed work intersects three research domains; public opinion polling, electronic voting (e-voting) systems, and blockchain technology. Each of these areas presents unique developments and ongoing challenges. In the following sections, we review relevant literature pertaining to each domain.

\subsection{Public Opinion Polling}
Public opinion polling has experienced substantial transformations over the past two decades, largely due to technological advancements and changing respondent behaviors. According to a 2023 Pew Research Center study, 61\% of U.S. national pollsters altered their polling methodologies between 2016 and 2022, reflecting a shift away from traditional methods \cite{kennedy2023public}. The decline of live telephone polling, which was once the dominant approach, has been driven by increasing costs and declining response rates. In response, many pollsters have turned to online surveys, probability-based panels, and Address-Based Sampling (ABS) to improve the representativeness and reliability of their data. These changes aim to modernize polling, improve data quality, and reduce costs. However, their effectiveness in predicting elections remains uncertain, especially given polling errors in 2016 and 2020 \cite{kennedy2018evaluation}. Ongoing research continues to assess the impact of these changes on the overall accuracy and credibility of public opinion measurements \cite{arletti2025making}.

Cerina et al. \cite{cerina2023artificially, cerina2025possum} examine how AI can support public opinion research by combining traditional surveys with large language models (LLMs) that extract structured, survey-like data from social media to improve representativeness, frequency, and cost-efficiency. In contrast, Boelaert \cite{boelaert2025machine} argues that LLMs often produce biased, low-variance responses that vary across topics, limiting their usefulness as substitutes for human respondents.

\subsection{e-Voting systems}
e-Voting systems have evolved over decades by adopting emerging technologies, with real implementations in countries like Estonia and Switzerland \cite{gerlach2009three, mpekoa2017voting}. Li et al. \cite{li2021publicly} address the lack of public traceability in anonymous authentication for e-voting by proposing Traceable Attribute-Based Anonymous Authentication (TABAA), which ensures anonymity, access control, and accountability without trusted third parties. Using zero-knowledge proofs and attribute-based credentials, it enables reusable, unlinkable authentication while preventing double voting through blockchain implementation for decentralized, transparent, and tamper-proof voting. Agrawal et al. \cite{agrawal2024publicly} introduce a system maintaining voter list privacy while enabling public audits to prevent fraud, allowing voters to verify their listing and auditors to detect fake participation or unfair removals without exposing personal data through cryptographic methods ensuring security and fairness.

\subsection{Blockchain Systems}
Since its introduction in the Bitcoin white paper \cite{nakamoto2008peer}, blockchain has attracted great interest for building decentralized crowdsourcing systems. Yao et al. \cite{yao2024distributed} propose a decentralized self-tallying e-voting protocol on Ethereum, combining zero-knowledge proofs and homomorphic encryption to ensure secrecy while enabling public ballot verification and result computation.

Muth et al. \cite{muth2023tornado} present Tornado Vote, a DApp for anonymous, fair voting on Ethereum, inspired by Tornado Cash to balance transparency and voter anonymity. Their protocol handles about 10,000 votes in two hours under optimal conditions. However, despite decentralization, current approaches still rely on a central authority for eligibility verification via government-issued IDs and PKI, limiting full decentralization.

\subsection{Filling the Gaps}
Our goal, in this work, is to address the limitations in the existing polling approaches by combining these three areas.  While modern polling approaches, like online surveys, can attract more responses due to ease of use and perhaps privacy preservation \cite{keeter2023publicopinion}, \cite{kennedy2023public}, that opens the door to fraudulent participation leading to wrong or fabricated results. In this work, we address the issue of declining response by harnessing the power of social influence while filtering ineligible responses using AI-assisted graph analysis.

e-Voting systems, including blockchain implementations \cite{9299581}, facilitate secure decentralized voting \cite{Ohize2024}. Due to high-impact results, e-voting requires strict eligibility verification, necessitating central authorities that undermine decentralization and privacy. Classical e-voting characteristics include anti-coercion and single participation \cite{gibson2016review}. While not implementing e-voting, we view polling as a relaxed voting version. Our proposed system replaces centralized eligibility checks with AI-assisted filtering, aiming for bounded error margins acceptable in polling domains. We allow multiple participations by design, which fights coercion when time-tracked. Future research will implement a prototype using modified blockchain technology for our application.



\begin{table*}[bht]
    \centering
    \renewcommand{\arraystretch}{1.2} 
    \begin{tabular}{|m{4cm}|m{12cm}|}
        \hline
        \textbf{Term} & \textbf{Definition} \\
        \hline
        \textbf{Eligible Node} & A participant that meets the predefined criteria for data collection.\\ 
        \hline
        \textbf{Ballot Unit} & A simulated unit that holds participants' responses. Each root node releases a predefined number of ballot units, which are distributed through the social graph. As nodes forward the poll their connections, a copy of a ballot unit is sent to the next participants. So, the number of these units increases as they propagate down the graph. For safety, each copy of a ballot Unit has a maximum capacity of 7 participations, while the tree-like propagation introduces a level of redundancy.\\
        \hline
        \textbf{Eligibility Ratio} & Represents the proportion of nodes that meet the criteria required to participate in the data collection process. \\
        \hline
        \textbf{Root Nodes} & Participants that serve as starting nodes in the dissemination graph and ignite information flows.\\ 
        \hline
        \textbf{Honest Nodes} & Participants who follow the voting rules and participate without malicious intent.\\
        \hline
        \textbf{Eligible Forwarding} & Sending the Ballot Unit to an eligible neighbor node so it can participate. \\
        \hline
        \textbf{Coverage} & Percentage of participants that received at least a single ballot unit relative to the total number of nodes in the graph. \\
        \hline
    \end{tabular}
    \caption{Key terms and definitions in the participation dissemination process}
    \label{tab:voting_definitions}
\end{table*}

\section{Proposed System} \label{sec:3}
To address our research questions, we imagine a scenario where a group of people characterized by a simple eligibility criterion is to be addressed by a poll. Instead of following the standard way of asking people to directly participate through mail or online portals, we leverage the power of social connections to disseminate the poll request. Specifically, we propose creating an induced social interaction by prompting the peers of a social network to participate and forward to others from their network of peers in a form similar to snowball sampling \cite{parker2019snowball}. Instead of having a central authority verify the eligibility criterion, we ask people to forward only to those who are eligible to participate where a receiver does not know the forwarder. The forwarding process follows a simple eligibility criterion that's easy to understand and comprehend among average people such as a specific age group, country, or district (similar to how real-world participation frameworks often apply). Using such a basic criterion enables easy and practical implementation while maintaining control over participant selection. It also keeps a low participation barrier by avoiding overly complex constraints that could cause confusion or restrict natural interaction dynamics. Naturally, some participants will commit to the eligibility request and some will not, depending on the level of honest they observe. 

\begin{figure*}
    \centering
    \begin{subfigure}{0.59\columnwidth}
        \centering
        \includegraphics[width=\linewidth]{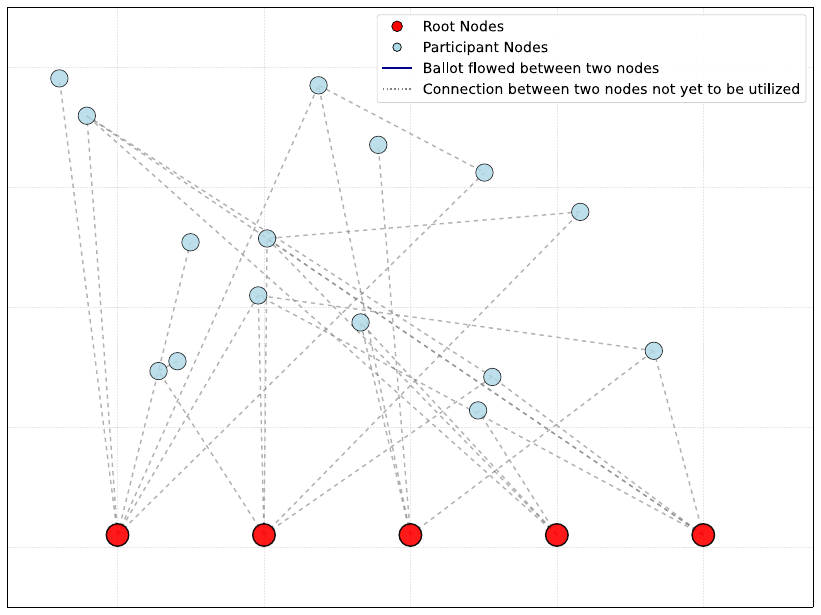}
        \caption{}
    \end{subfigure}
    \hfill
    \begin{subfigure}{0.59\columnwidth}
        \centering
        \includegraphics[width=\linewidth]{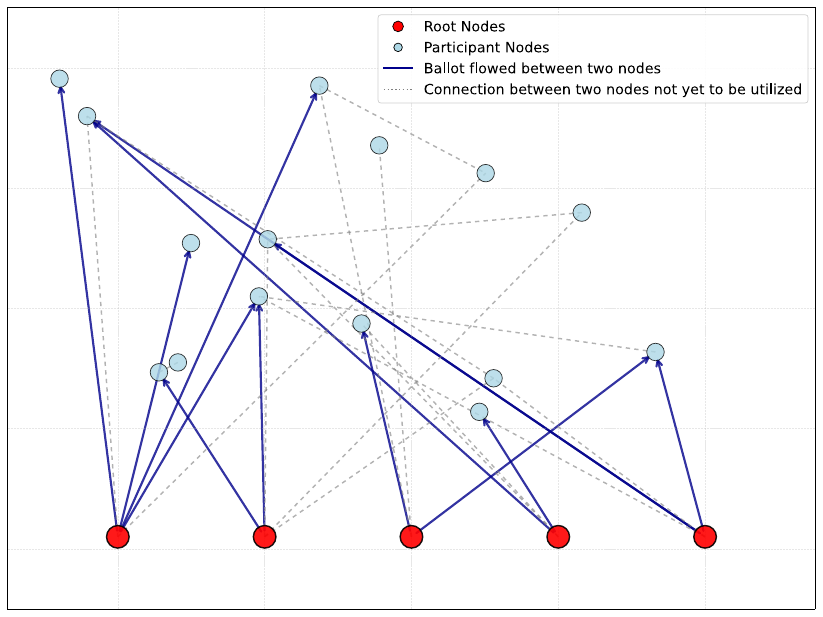}
        \caption{}
    \end{subfigure}
    \hfill
    \begin{subfigure}{0.59\columnwidth}
        \centering
        \includegraphics[width=\linewidth]{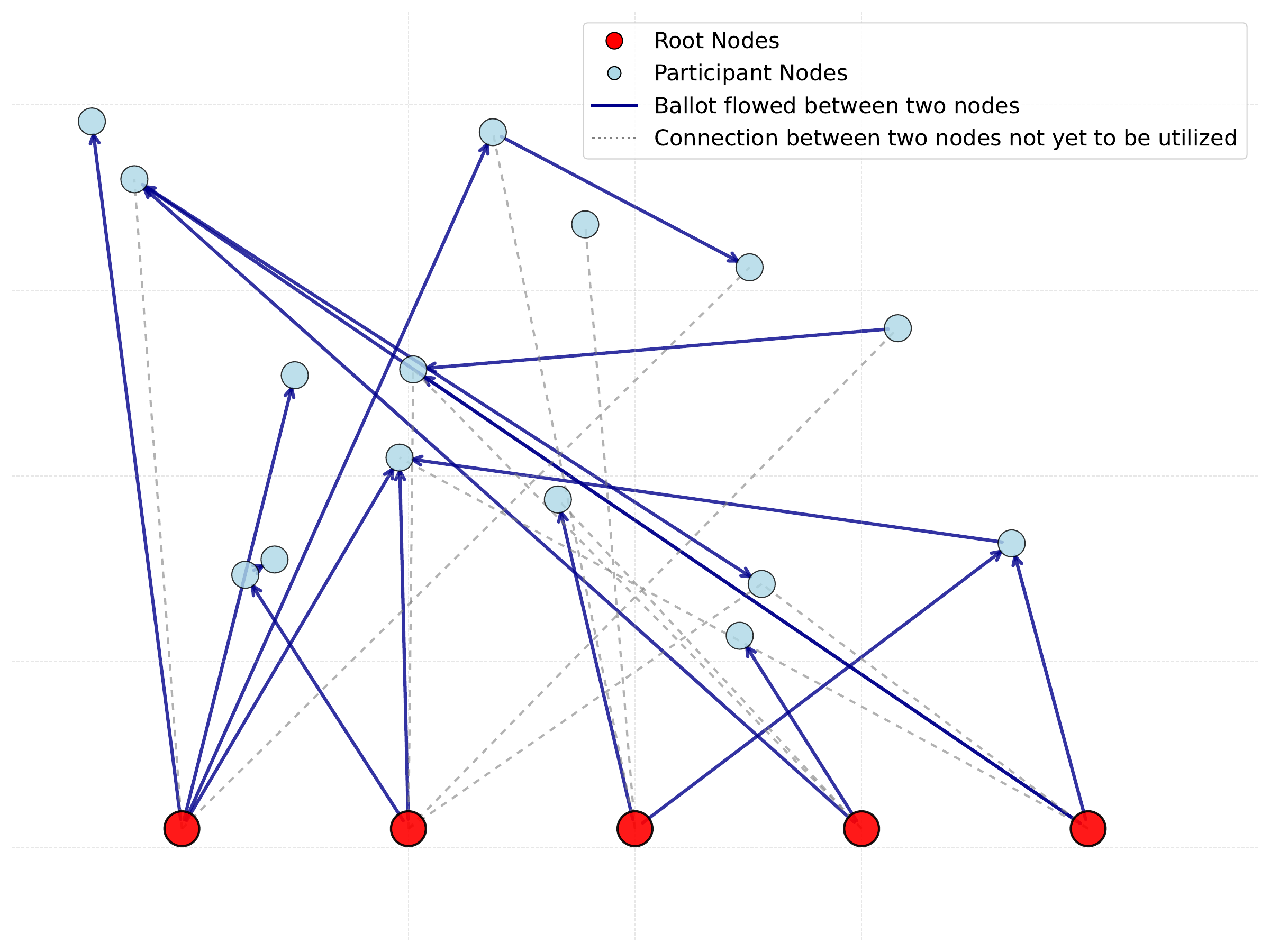}
        \caption{}
    \end{subfigure}
    \caption{Evolution of the dissemination graph over time (a $\rightarrow$ c)}
    \label{fig:dissemination}
\end{figure*}

We simulate simple honest behavior as it will determine the data flow in the social network. To motivate more honesty, we start the participation flow from credible members of the social network or the society representing key figures in a community (we call them \textit{root nodes} in our experiments and results section). Due to their distinction, these key participants are expected to pass more traffic than an average participant. We assume a certain percentage of honest participants (randomly selected) at the beginning of the simulation where honest participants are more likely to commit to the eligibility criterion than dishonest ones. Like in cryptocurrency systems, we assume there is a way to securely and anonymously identify unique participants, hence allow them to respond to the poll and/or forward to others multiple times as they wish counting only the last participation as the official response. Multi-chance participation/forwarding widens the dissemination scale, increases social interaction, and creates a level of redundancy that we hope can all help differentiate eligible from ineligible participants through their response/forwarding behavior. It also satisfies the anti-coercion requirement in e-voting systems\cite{gibson2016review}. We use a cap to the length of a forwarding sequence to allow for new flows to stream through the social network for enhanced exploration. 
The participants submit their responses to an advisory entity (that could also be decentralized like in blockchain applications \cite{peterson2019augur}), regardless of their decision to forward to others. We assume information about the forwarding flow can be anonymously and securely encoded into the response where a graph of the whole social interactions can be constructed from the submitted responses without revealing the identity of the respondents, called a \textit{a dissemination graph}. This graph is typically a subset of the original social graph but with more information about how it was created. We use machine learning to allow a model to learn the ineligible participants from the social interaction information embedded into the dissemination graph. Figure \ref{fig:dissemination} shows how the dissemination graph is generated over time, with ballot dissemination initiated from root nodes and shaped by network structure and participant behavior.


Modeling social interaction in this experiment was challenging. First, most available social network datasets are highly anonymized, making it difficult to find one suitable for use as a participation criterion. Second, although studies exist on honest social behavior \cite{ayal2012honest}, translating their findings into quantifiable parameters for computer simulations proved challenging. Therefore, we had to use probability distributions to model these behaviors as close as we can. We also needed to make more design choices that are plausible from a practical as well as behavioral viewpoint. For example, how many times a user would participate and forward the poll request before feeling bored and how many connections a user would forward to. These and other questions would be better learned from an actual experiment, which we intend to do in a later stage of this research.



\section{Key System Configurations} \label{sec:4}
In this section, we define key system terms and describe their configurations. These terms are summarized in Table~\ref{tab:voting_definitions} and we provide more details below.

\subsection{Honest Behavior}
Unlike structural properties in a network, honesty is inherently person-aware and a self-trait, making it difficult to quantify using simple metrics \cite{ashton2014hexaco, verschuere2018registered}. While people can exhibit a diverse spectrum of honest behavior, we classify participants as either honest or dishonest for our initial investigation. For instance, if the eligibility criterion for participation is a minimum age of 21, an honest participant — whether eligible or ineligible — strictly adheres to the rule and forwards the poll to its connections who are more than 21 years of age. Conversely, a dishonest participant is willing to engage in a malicious or manipulative behavior by forwarding to both eligible and ineligible peers. At the beginning of our simulations, we set an honest ratio and randomly assign honesty to the graph nodes accordingly.

\subsection{Root Nodes}
In real-world scenarios root nodes should ideally be honest and trusted individuals with high reputation and are well-connected across their larger network. To demonstrate this in our work, we used centrality measures to approximate this selection. Specifically, we use the Betweenness centrality measure \cite{freeman1977set, bavelas1948mathematical} as our metric for selecting nodes, as it identifies graph vertices that act as critical bridges between different parts of the graph. Nodes with higher Betweenness centrality lie on a larger number of shortest paths, making them influential in facilitating information flow across the network. By leveraging these high-centrality nodes, we enhance the poll propagation, ensuring broader participation by reaching diverse parts of a targeted community. Due to their critical role in the information dissemination process, all of the selected root nodes are chosen from the honest and eligible node sets. We assume this description and behavior of the root nodes matches society key figures or influencers in a real-life scenario. A careful selection of these root nodes along with detecting ineligible participants (a main contribution of this paper) helps alleviate possible biases of the snowball-like dissemination process.
\subsection{Participation Behavior}

We designed our experimental setup to closely mirror real-world scenarios, considering the diverse behaviors of individuals in anonymous participation rounds. In a real world scenario, participants will exhibit varying appetite for filling a poll and/or forwarding to others. To simulate this behavior, we use the node degree centrality measure to set the number of participations for each node separately. We sample this number from an exponential distribution such that $90\%$ of the participation count ($\alpha=0.9$) lie below $10\%$ of a node’s degree ($\beta=0.1$), as shown in Equation~\ref{equ:particip}, with a minimum of 1 participation. We compute the rate parameter $\lambda$ parameterized by the node's degree $d$, where $\beta d$ represents $90\%$ of the node's degree. By tying the participation limit to node degree in this way, we aim to mimic the inclination of well-connected people to share more. Additionally, in our setting, root nodes are granted an additional fixed number of participation opportunities to aid their role in the process. This allocation is directly tied to the initial count of ballot units assigned to each root node, facilitating effective dissemination throughout the network. We set these extra participation opportunities to 30.



\begin{equation}
\label{equ:particip}
\begin{aligned}
& \text{\textit{participation count}} \sim \max\!\left(\lfloor \mathrm{Exp}(\lambda(d)) \rfloor, \, 1\right) \\
& \text{where } \lambda(d) = \frac{-\ln(1 - \alpha)}{\beta d}, \alpha = 0.9, \; \beta = 0.1
\qquad 
\end{aligned}
\end{equation}








\subsection{Eligibility Criteria}
Participation eligibility is central to developing our system. We adopt a flexible and adaptable criterion that supports diverse use cases across different scenarios. Eligibility can be defined by factors such as age, country of residence, or affiliation with a particular group, reflecting real-world practices where individuals must meet minimum requirements to participate. For instance, in voting processes, citizens are typically required to satisfy conditions like age and residency to qualify.

\subsection{Ballot Units Forwarding}
Each non-root node that received a ballot unit - whether it participated or not - can forward it to one or more of its social connections. To ensure a realistic and fair selection, we employ the round-robin technique to determine the next connection to receive the ballot unit. This approach ensures balanced distribution and helps reduce bias in the forwarding process. 
The forwarding behavior differs between honest and dishonest participants. Regardless of its eligibility status, an honest participant will forward only to eligible connections. A dishonest participants, though, may forward to both eligible and ineligible connections. To simulate this behavior we follow a probabilistic decision-making process. Each time a dishonest participant encounters a ballot unit, it samples a value from a uniform distribution. Using a randomized threshold decision rule, the node compares this value against a probability to decide whether to behave legitimately or act maliciously when forwarding. Eligible nodes refrain from sending if they have no eligible neighbors, whereas ineligible nodes send to their eligible neighbors only when they lack ineligible ones.

This approach is justified by previous research \cite{gino2009contagion, ayal2012honest}, which highlights key observations about the behavior of malicious participants. Studies suggest that dishonest actors tend to be strategic and cautious, aiming to conceal their fraudulent activities to avoid detection and protect their reputation. Additionally, they often strive to maintain a positive self-image, both publicly and privately. To minimize suspicion, these actors do not engage in continuous fraudulent behavior; instead, they mix malicious actions with legitimate ones. This strategic alternation ensures that their overall conduct appears partially compliant rather than overtly fraudulent, reducing the likelihood of exposure.

\begin{table}[h]
    \centering
    \caption{Summary of datasets used}
    \resizebox{\columnwidth}{!}{%
    \begin{tabular}{lccc}
        \toprule
        \textbf{Dataset} & \textbf{Source} & \textbf{Nodes} & \textbf{Edges} \\
        \midrule
        Last.fm Multigraph & UCI Networking Group & 465,166 & 4,120,950 \\
        Musae-Twitch (DE)  & SNAP & 9,498  & 153,138 \\
        \bottomrule
    \end{tabular}%
    }
    \label{tab:datasets}
\end{table}

\section{Experiments}\label{sec:5}
We conducted simulations on multiple datasets to evaluate model performance in predicting participant eligibility under varying community conditions, including the proportions of eligible users, root nodes, and honest participants.
\subsection{Datasets}
While many datasets exhibit the social-network structure we need for our experiments, we needed datasets that meets the following criteria to fit our empirical study:

\begin{enumerate}
    \item The dataset has to include individual-based features to be used as eligibility criterion. These features either can be categorical or bear multiple values allowing us to segment the social network based on different values of this feature(s). For example, a feature like age group can be useful, as it enables categorization into predefined ranges (e.g., under 18, 18–30, 31–50, and 50+), facilitating meaningful analysis. However, descriptive features like objectives, self-reflections would not be effective. It was challenging to find datasets with this criteria because most of the available social networks datasets are highly anonymized.
    \item The dataset should contain edges that represent real social connections to allow us to study authentic human behavior rather than artificial or random connections.
\end{enumerate}

We conducted our experiments using two datasets, 1 from UCI Networking Group and 2 datasets from SNAP \cite{snapnets}, while exploring diverse domains for the eligibility criteria. These simple characteristics of these datasets are summarized in Table~\ref{tab:datasets}, and in details as follows:



\begin{enumerate}
    \item Last.fm Multigraph Dataset \cite{lastfm}: This dataset is an Internet website for music with social networking features. For this dataset, we used the age attribute of each participant as the primary eligibility criterion. We constructed the edges based solely on the \texttt{uname\_friends} relationship.
    The dataset contained a large number of connected components, exceeding 50,000, with the majority being outliers. For our analysis, we selected the largest connected component (465,166 nodes), as it provided the most meaningful structure. The second-largest component, in comparison, had a size of only seven nodes.
    
    \item Musae-Twitch Dataset \cite{twitch}: This dataset consists of Twitch user-user networks from May 2018, representing gamers who stream in a specific language. Nodes represent users, edges represent friendships, and features include games played, location, and streaming habits. We used the minimum total channel views reported for a user as the eligibility criterion. We used the Germany dataset as it had the highest number of nodes compared to other countries.  
\end{enumerate}

\subsection{Experiment Setup}

The approach we employed to capture the characteristics of social interactions within the network was based on graph node embeddings, which encode the relationships between nodes and the overall graph structure. We generated both static and dynamic node embeddings. Static embeddings were generated using Node2Vec \cite{grover2016node2vec}, while dynamic embeddings were learned using GraphSAGE \cite{hamilton2017inductive} with an embedding vector length of 32, which were jointly refined during classification. GraphSAGE was used also as a classifier to predict the node eligibility given the combined static and dynamic node embeddings. Both static and dynamic embeddings were concatenated for the GraphSAGE classifier.

To optimize performance, we tuned the Node2Vec parameters, including embedding vector size, random walk length, and number of walks. Since the two datasets exhibited structural differences, we used dataset-specific configurations for random walk length and number of walks. In both datasets, the embedding dimension was set to 64. For the Twitch dataset, we used a walk length of 32 with 50 walks per node, whereas for the Last.fm dataset, we applied a walk length of 50 with 30 walks per node.

We further experimented with the proportion of root nodes, testing values of 1\% and 5\%, and capped it at 5\% to constrain the number of influencers. Similarly, we varied the proportion of honest nodes, using values of 60\% and 80\%. In addition, we evaluated four different eligibility ratios per dataset to investigate how model performance varied across configurations. Establishing comparable eligibility ratios across datasets was essential for fair evaluation. However, this proved challenging because the relationship between eligibility criteria and the resulting eligibility ratio is nonlinear. To address this, we tested multiple configurations and selected values that yielded ratios that were close, though not identical, across the two datasets, as summarized in Table~\ref{tab:eligibility}.

\begin{table}
    \centering
    \caption{Eligibility ratios used across datasets}
    \label{tab:eligibility}
    \renewcommand{\arraystretch}{1.2}
    \resizebox{\columnwidth}{!}{%
    \begin{tabular}{lcccc}
        \toprule
        \textbf{Dataset} & \textbf{Eligibility 1} & \textbf{Eligibility 2} & \textbf{Eligibility 3} & \textbf{Eligibility 4} \\
        \midrule
        Last.FM  & 86.66\% & 56.87\% & 38.67\% & 26.2\% \\
        Twitch   & 88.67\% & 55.4\%  & 38.69\% & 24.86\% \\
        \bottomrule
    \end{tabular}%
    }
\end{table}



\section{Results} \label{sec:6}

\subsection{Prediction Power}
\begin{figure}
    \centering
    \begin{subfigure}{0.48\columnwidth}
        \centering
        \includegraphics[width=\linewidth]{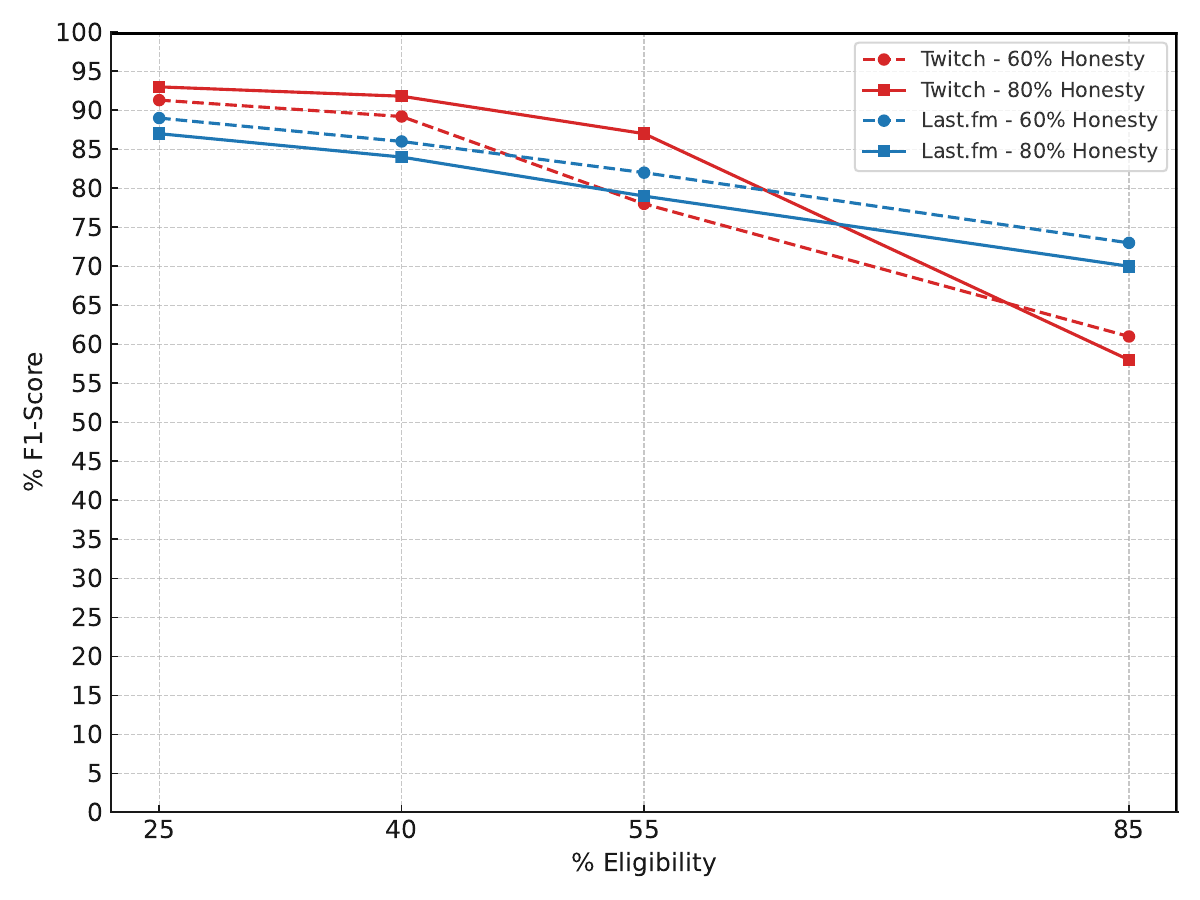}
        \caption{1\% root nodes}
        \label{fig:F1score-1perc-root-nodes}
    \end{subfigure}
    \hfill
    \begin{subfigure}{0.48\columnwidth}
        \centering
        \includegraphics[width=\linewidth]{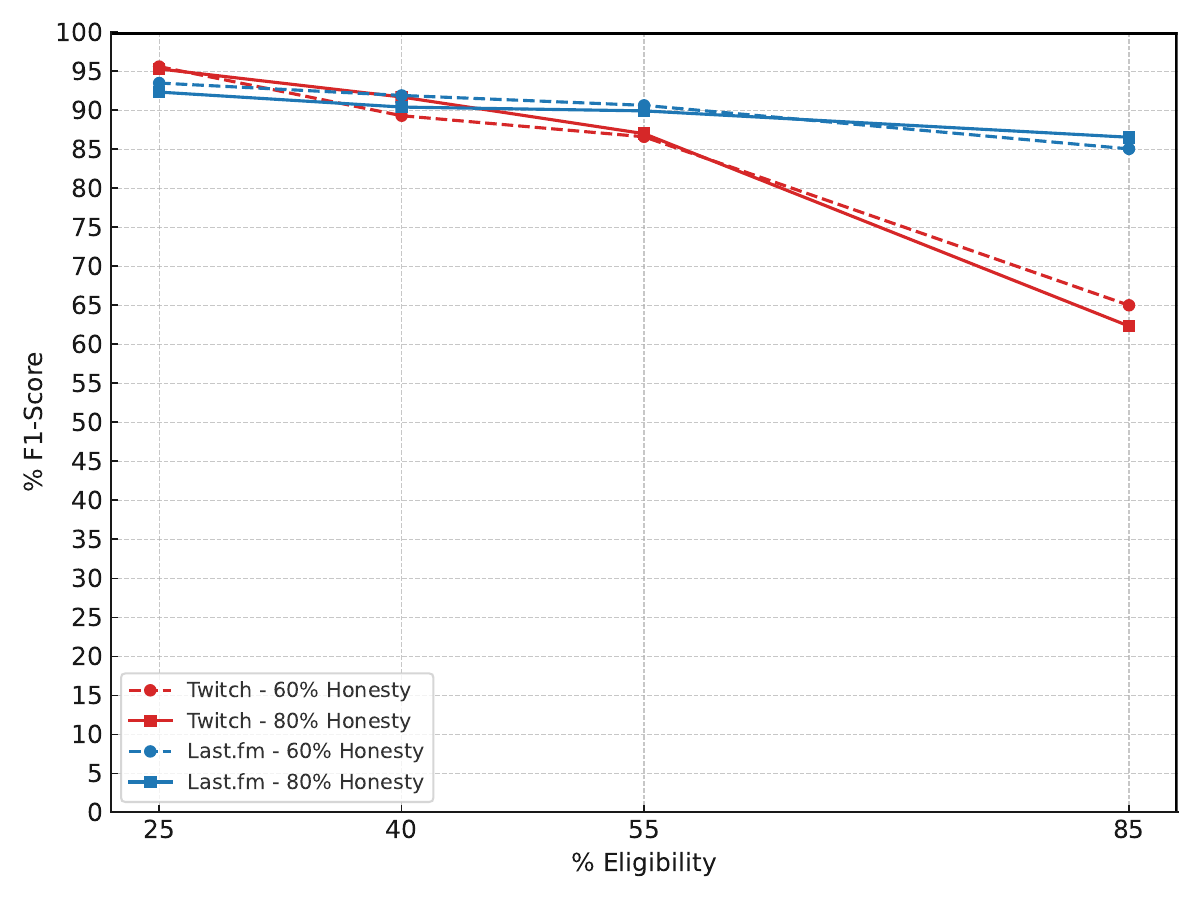}
        \caption{5\% root nodes}
        \label{fig:F1score-5perc-root-nodes}
    \end{subfigure}
    \caption{Comparison of F1-scores for two datasets under varying eligibility ratios, root coverage, and participant honesty levels.}
    \label{fig:F1-score}
\end{figure}

\begin{figure}
    \centering
    \begin{subfigure}{0.48\columnwidth}
        \centering
        \includegraphics[width=\linewidth]{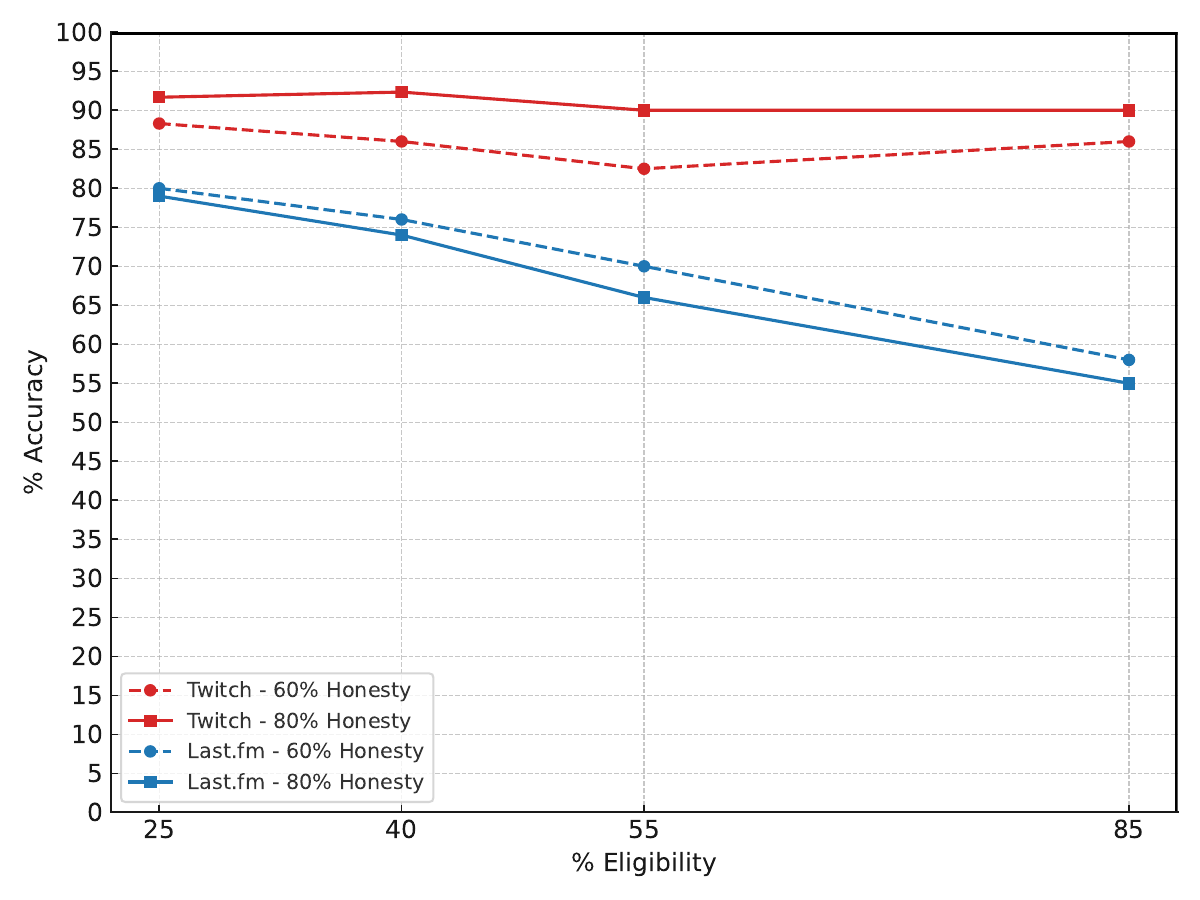}
        \caption{1\% root nodes}
        \label{fig:accuracy-1perc-root-nodes}
    \end{subfigure}
    \hfill
    \begin{subfigure}{0.48\columnwidth}
        \centering
        \includegraphics[width=\linewidth]{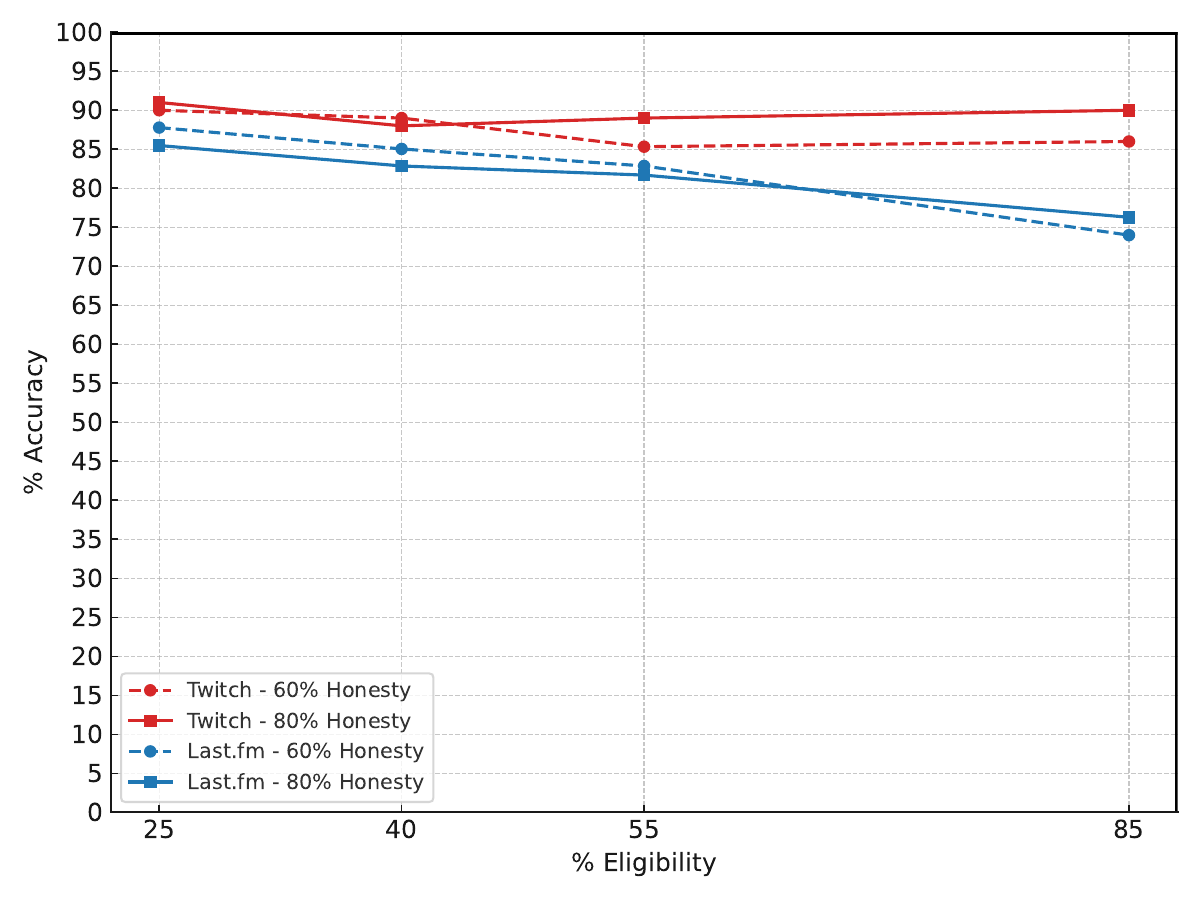}
        \caption{5\% root nodes}
        \label{fig:accuracy-5perc-root-nodes}
    \end{subfigure}
    \caption{Accuracy for the two dataset against different levels of eligibility ratios across different root node levels and different levels of participant honesty.}
    \label{fig:accuracy}
\end{figure}

Figures \ref{fig:F1-score} and \ref{fig:accuracy} show the obtained F1-score and accuracy using the graph node embeddings against different levels of eligibility ratios for different levels of root nodes across the two different datasets. While the two datasets exhibit a similar trend, each have a different bracket of performance reflecting the impact of the graph structure that naturally changes across social networks. The figure also shows that better predictions can be obtained as the percentage of root nodes increases. This is because root nodes are honest participants who are more likely to forward to eligible participants.

\subsection{Coverage}
\begin{figure}
    \centering
    \begin{subfigure}{0.47\columnwidth}
        \centering
        \includegraphics[width=\linewidth]{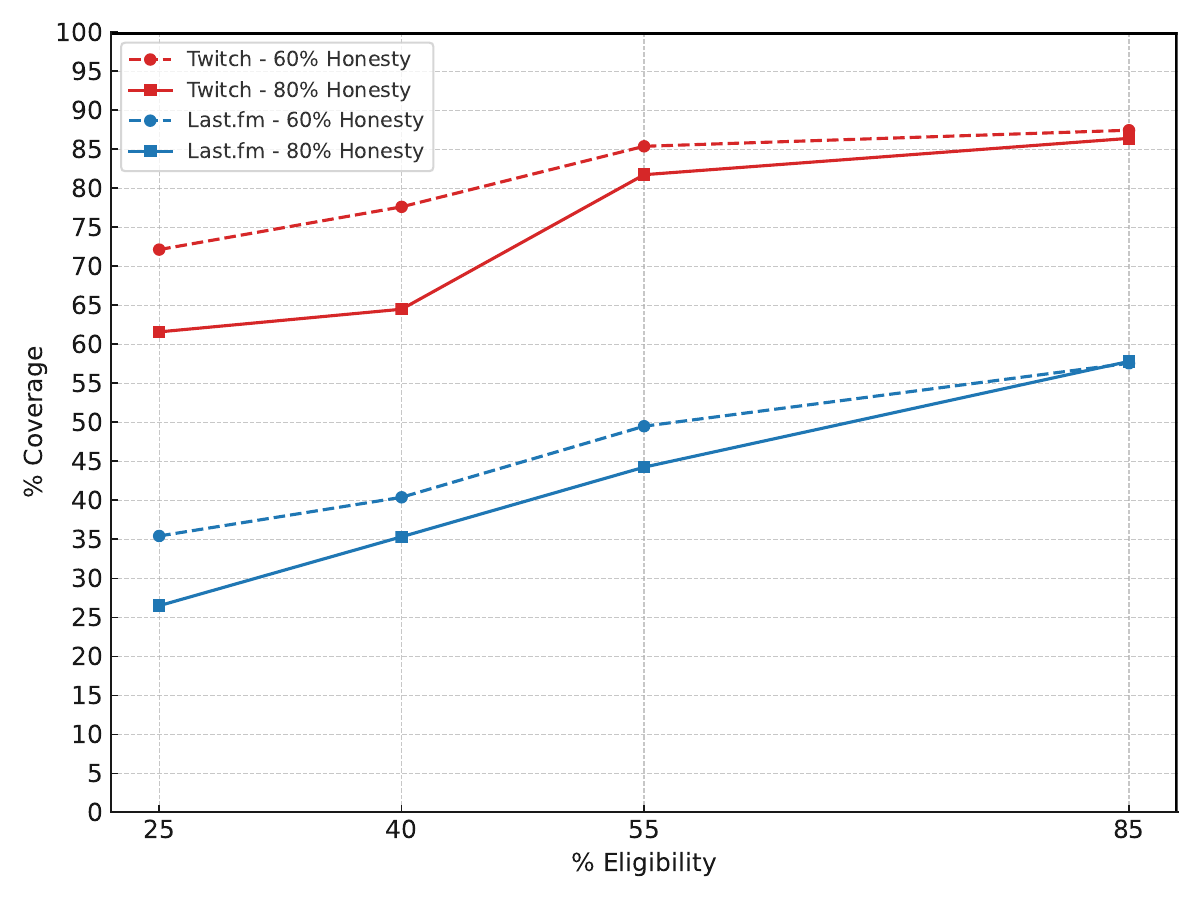}
        \caption{Coverage for 1\% root nodes}
    \end{subfigure}
    \hfill
    \begin{subfigure}{0.47\columnwidth}
        \centering
        \includegraphics[width=\linewidth]{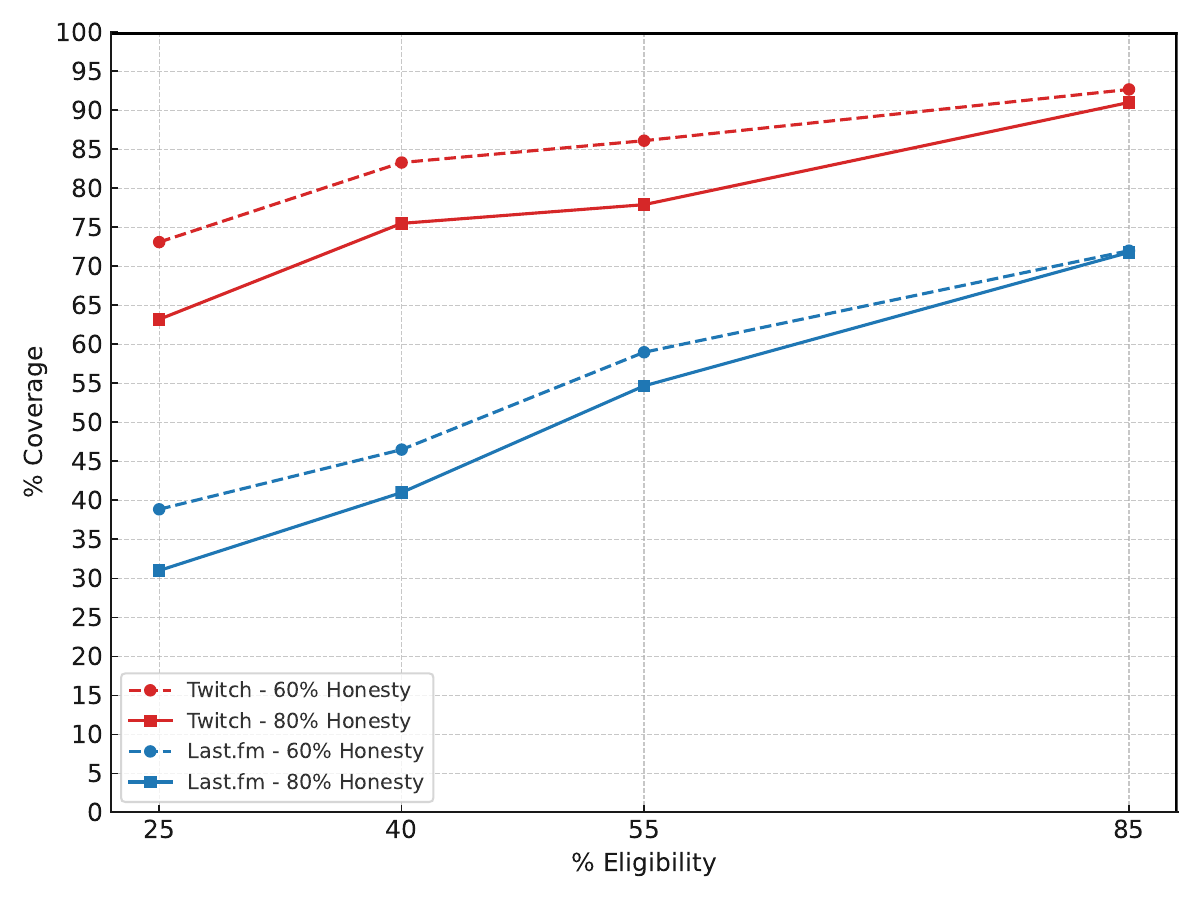}
        \caption{Coverage for 5\% root nodes}
    \end{subfigure}
    \caption{Coverage \% across 1\% and 5\% root nodes.}
    \label{fig:coverage}
\end{figure}

We study the coverage, as defined in Table \ref{tab:voting_definitions}, to inspect how much of the target population have received the poll request. We examined the two cases, 1\% root nodes and 5\% root nodes, for each dataset under the highest percentage of eligibility across different honesty levels (60\%, 80\%) as shown in figure~\ref{fig:coverage}. The low coverage in lower eligibility ratio scenarios is reasonable because ineligible nodes are naturally less likely to receive a participation ballot than in a higher illegibility ratio scenario given the same level of honesty. Figure~\ref{fig:coverage} also shows that the coverage in the Twitch dataset is significantly higher than in the Last.FM dataset. We attribute this to the fact that the nodes in the Twitch dataset have a higher average node degree compared to the Last.fm dataset. Specifically, the Twitch dataset has an average of 32 edges per node, while the Last.FM dataset has 17 edges per node. This affirms again the impact of the graph structure and connectivity on the sought results. We take this into consideration while tuning the system configurations in our ongoing work extension to develop a generalized model.

\subsection{Practicality and Applicability}
\begin{figure}
    \centering
    \begin{subfigure}{0.46\columnwidth}
        \centering
        \includegraphics[width=\linewidth]{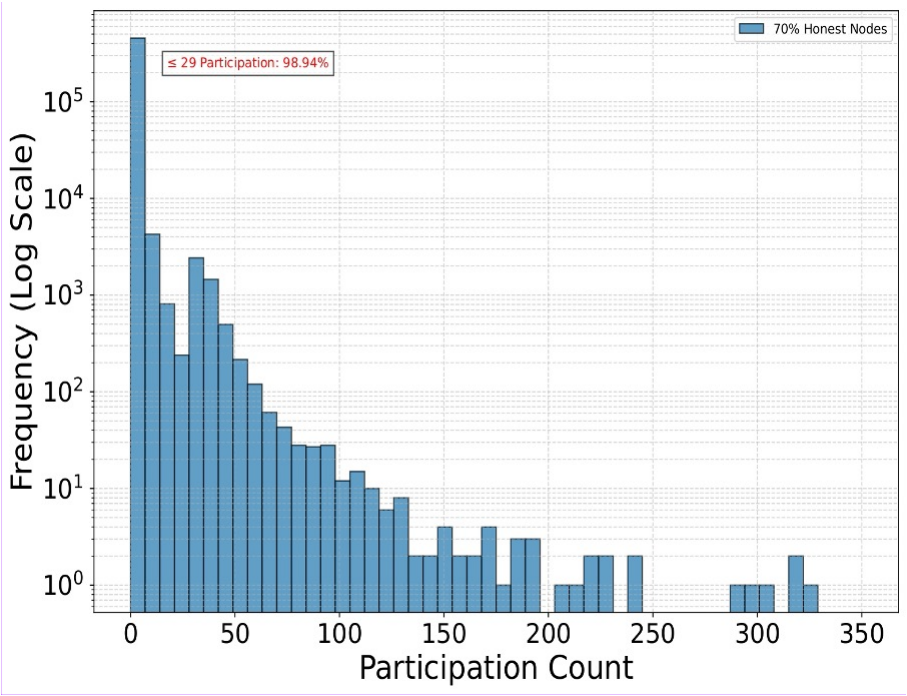}
        \caption{1\% root nodes}
        \label{fig:votes_hist_1_perc}
    \end{subfigure}
    \hfill
    \begin{subfigure}{0.46\columnwidth}
        \centering
        \includegraphics[width=\linewidth]{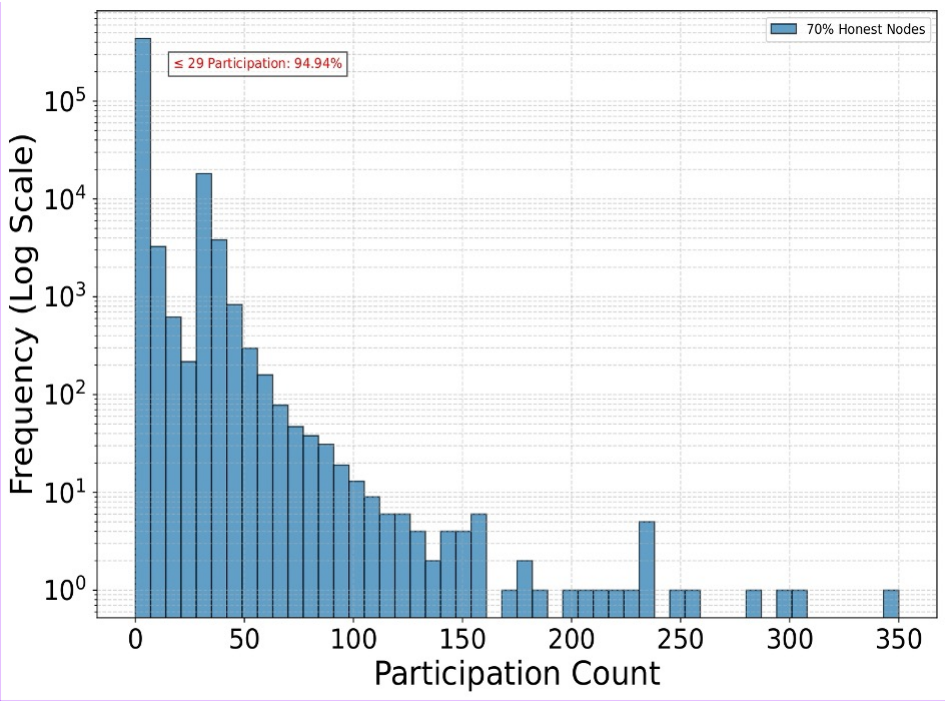}
        \caption{5\% root nodes}
        \label{fig:votes_hist_5_perc}
    \end{subfigure}
    \caption{Last.FM Dataset: Histogram for participation count at different root node levels at 70\% honest nodes.}
    \label{fig:last_fm_votes_hist}
\end{figure}

To assess our system’s practicality and applicability, we measured each participant’s participation frequency which is the number of ballots received and acted upon. Figure~\ref{fig:last_fm_votes_hist} (log scale) shows the distribution for 1\% and 5\% root nodes. Nearly all non-root nodes participated fewer than the 30 additional forwards allowed for roots, with most engaging far less. Although 30 participations is not excessive, reducing this load remains a goal for future work, and the results suggest our setup is feasible with reasonable interaction levels.

\begin{figure}
    \centering
    \begin{subfigure}{0.45\columnwidth}
        \centering
        \includegraphics[width=\linewidth]{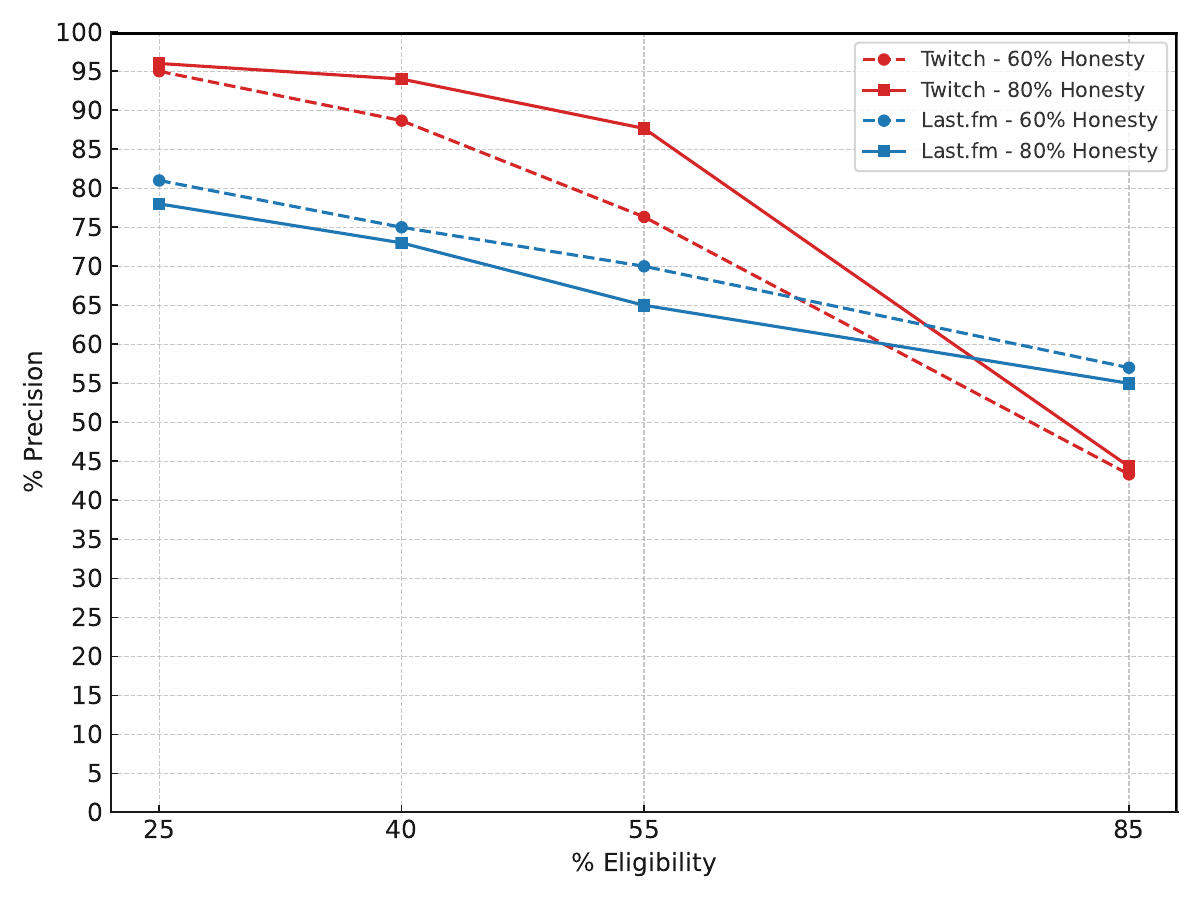}
    \end{subfigure}
    \hfill
    \begin{subfigure}{0.45\columnwidth}
        \centering
        \includegraphics[width=\linewidth]{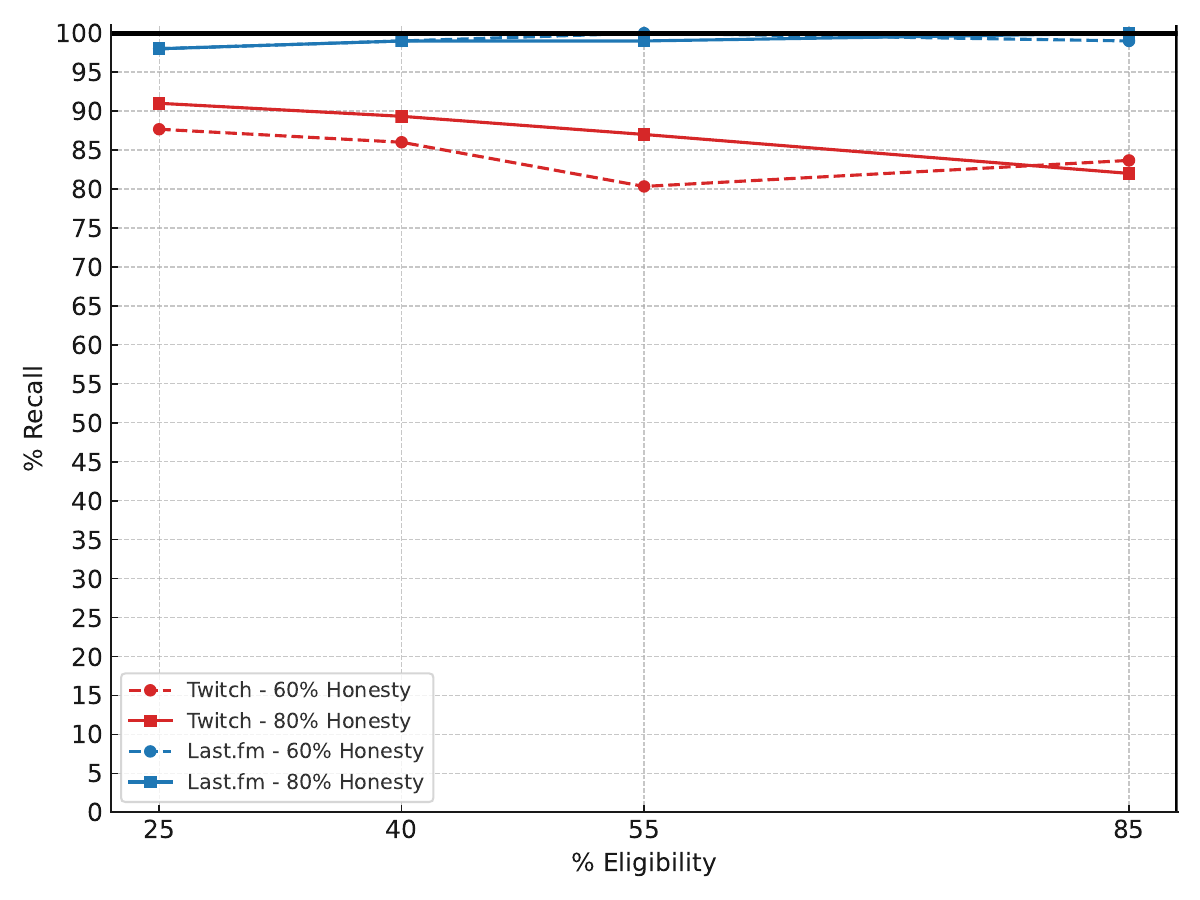}
    \end{subfigure}
    \caption{Precision and recall at 1\% root nodes}
    \label{fig:prec_recall}
\end{figure}

\begin{figure*}
    \centering
    \begin{subfigure}{0.59\columnwidth}
        \centering
        \includegraphics[width=\linewidth]{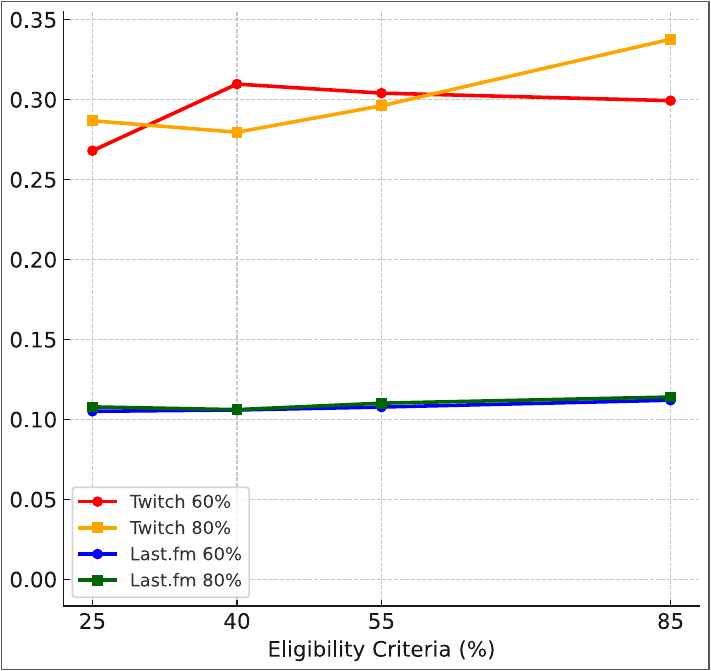}
        \caption{Clustering}
    \end{subfigure}
    \hfill
    \begin{subfigure}{0.59\columnwidth}
        \centering
        \includegraphics[width=\linewidth]{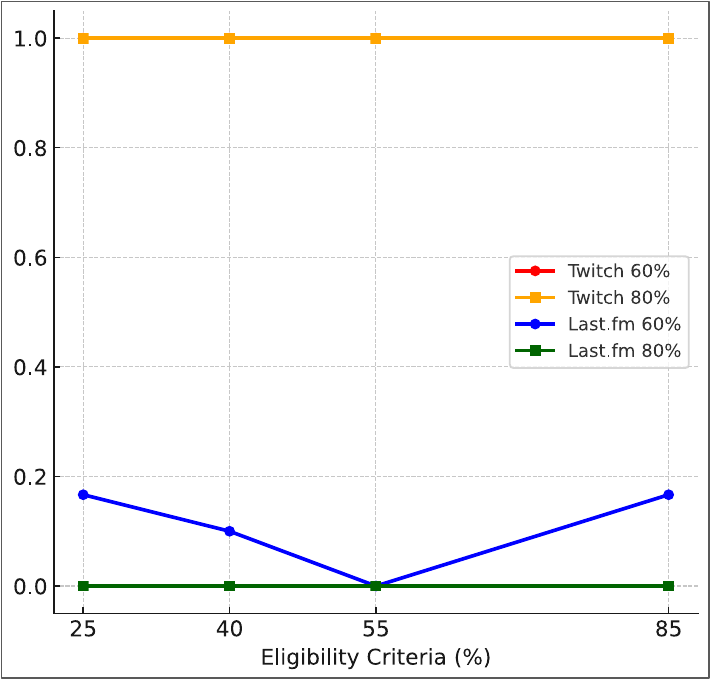}
        \caption{Rich Club Coefficient}
    \end{subfigure}
    \hfill
    \begin{subfigure}{0.59\columnwidth}
        \centering
        \includegraphics[width=\linewidth]{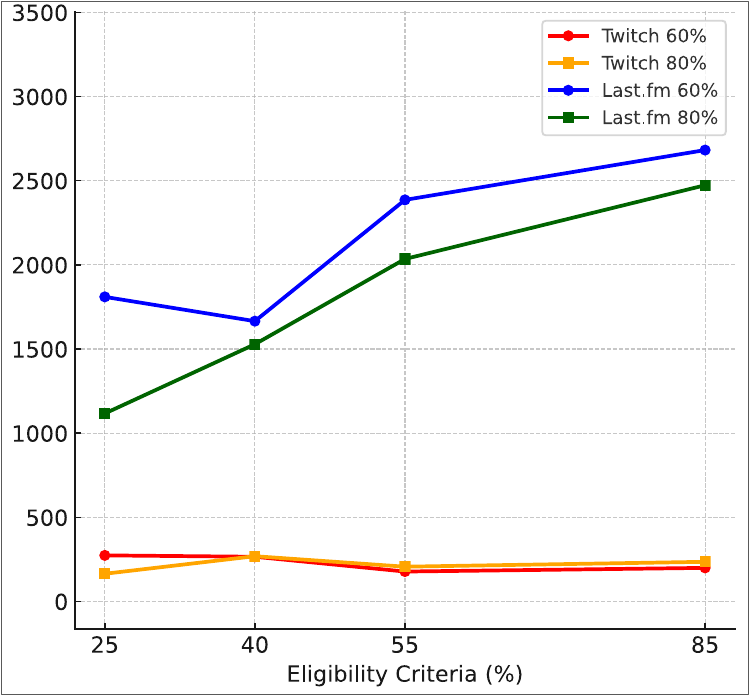}
        \caption{Community Size}
    \end{subfigure}
    \caption{Clustering, rich-club coefficient, and community size reveal complementary aspects of the network’s structure}
    \label{fig:network_struc}
\end{figure*}

\section{Discussion} \label{sec:7}
Our proposed system achieved over 80\% F1-score in detecting ineligible participants in the lower eligibility ratio division ($\leq 55\%$), with degrading performance in the upper eligibility ratio. This indicates that the model was able to learn and detect ineligible participants better in the lower eligibility ratio scenarios, which is reasonable as more ineligible participants are present in this lower division. On the other hand, a higher proportion of root nodes enhanced accuracy, reflecting their role as credible and trusted actors. Looking deeper, Figure \ref{fig:prec_recall} shows the values for precision and recall in the 1\% root node scenarios. We can observe that the recall was significantly higher than precision. This underscores the model’s ability to correctly identifying actual ineligible participants, which we think is more desired in our problem that improving the precision, if we had to choose. However, we aim to improve the precision as well as we further our research in this problem.

A close look at the dissemination graphs structure of the two social datasets reveals fundamental differences that can impact the participants’ behavior and consequently the prediction power of our model. We identify and plot three measures to compare the graph structures Figure \ref{fig:network_struc}:

\begin{itemize}
    \item \textbf{Clustering}\cite{watts1998collective}: Likelihood that a node’s neighbors are also connected. Higher means local "cliquishness”
    \item \textbf{Rich club coefficient} \cite{zhou2004rich}: Tendency of high-degree nodes to connect to each other. Higher means presence of elite core of interconnected hubs.
    \item \textbf{Community size} \cite{blondel2008fast}: Size of detected structural groups (via Louvain or label propagation). Higher variation/skew means network has diverse community structure
\end{itemize}

Figure \ref{fig:network_struc} shows that the Twitch community is more homophilous \cite{khanam2023homophily, bisgin2010investigating} than Last.fm, where higher clustering and rich club coefficients, and lower variation in community structure can be observed in the Twitch network. Therefore, we can expect dishonest behavior to be more odd, hence more observable, in Twitch than in Last.fm. This can explain the difference in prediction power between the two datasets. 

For example, the users in the Twitch network tend to form tight-knit groups, like circles of close friends where many people know each other. This is shown by the higher clustering score \cite{watts1998collective} $(\sim 0.29)$ compared to Last.FM dataset. So, unless there is a wide scale collusion within the tightly connected groups, a dishonest behavior (breaking common rules) could be easier to detect. In contrast, Last.fm has very low clustering $(\sim 0.10)$. This means users are more loosely connected,  more like distant acquaintances. Without strong local connections, it could be harder for a model to learn who’s behaving anomalously, so dishonest participations blend in more easily, making it harder for the model to detect malicious behavior. Contemplating these measures motivates us to improve our prediction models to be more robust against these graph structural differences.




\section{Conclusions and Future Work} \label{sec:8}
In this paper, we present an empirical study on how social interactions can enhance the credibility of polls. We use the social interaction to spread poll requests, and apply AI with node embeddings to analyze the resulting dissemination graph, enabling the detection of ineligible participants. Using two real-world datasets, our models achieved promising accuracy, highlighting a new direction for improving polling results with broader participation and improved reliability. 
Although we approximated human behavior in our simulations, relying on synthetic social data is a limitation since it omits real social and psychological dynamics. We plan to expand this study with real life experiments to refine our findings. We also plan to  develop generalizable models that is robust against different graph structures.

\bibliographystyle{IEEEtran}
\bibliography{references}

\end{document}